\documentclass[12pt,a4paper]{article}
\usepackage[T2A]{fontenc}
\usepackage{amsfonts}
\usepackage{amsmath}
\usepackage{amssymb}
\usepackage[english]{babel}
\usepackage{graphicx}
\begin{document}
\title{The dynamics of binary alternatives for a discrete pregeometry} 

\author{Alexey L. Krugly\thanks{Scientific Research Institute for System Analysis of the Russian Academy of Science, 117218, Nahimovskiy pr., 36, k. 1, Moscow, Russia; akrugly@mail.ru.}}
\date{} \maketitle
\begin{abstract}
A particular case of a causal set is considered that is a directed dyadic acyclic graph. This is a model of a discrete pregeometry on a microscopic scale. The dynamics is a stochastic sequential growth of the graph. New vertexes of the graph are added one by one. The probability of each step depends on the structure of existed graph. The particular case of dynamics is based on binary alternatives. Each directed path is considered as a sequence of outcomes of binary alternatives. The probabilities of a stochastic sequential growth are functions of these paths. The goal is to describe physical objects as some self-organized structures of the graph. A problem to find self-organized structures is discussed. 

\bigskip\noindent\textbf{Keywords:} causal set, random graph, self-organization.

\noindent\textbf{PACS:} 04.60.Nc
\end{abstract}

\section{Introduction}
\begin{figure}[ht]
	\centering	
		\includegraphics[trim=8cm 18cm 8cm 7cm]{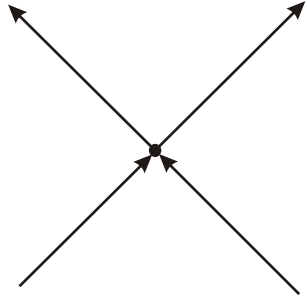}
	\caption{An x-structure.}
	\label{fig:fig1}
\end{figure}

Consider a particular model of a discrete pregeometry. This is a directed dyadic acyclic graph. The edges are directed. Each vertex possesses two incident incoming edges and two incident outgoing edges. A vertex with incident edges is called an x-structure (Fig.\ \ref{fig:fig1}).
The model was introduced by D. Finkelstein in 1988 \cite{Fink88}. The acyclic graph means that there is not a directed loop. In this paper only such graphs are considered. Then they are called graphs for simplicity.

This model is the particular case of a causal set. A causal set is a pair ($\mathcal{C}$, $\prec$), where $\mathcal{C}$ is a set and $\prec$ is a binary relation on $\mathcal{C}$ satisfying the following properties ($x,\ y,\ z$ are general elements of $\mathcal{C}$):
\begin{equation}
\label{eq:I1.1} x\nprec x\qquad\textrm{(irreflexivity),}
\end{equation}
\begin{equation}
\label{eq:I1.2} \{x\mid(x\prec y)\wedge(y\prec x)\}=\emptyset \qquad \textrm{(acyclicity),}
\end{equation}
\begin{equation}
\label{eq:I1.3} (x\prec y)\wedge(y\prec z)\Rightarrow(x\prec z)\qquad \textrm{(transitivity),}
\end{equation}
\begin{equation}
\label{eq:I1.4} \mid\mathcal{A}(x, y)\mid<\infty\qquad\textrm{(local finiteness),}
\end{equation}
$$
\textrm{where\ }\mathcal{A}(x,\ y)=\{z\mid x\prec z \prec y\}\textrm{.}
$$
The first three properties are irreflexivity,  acyclicity, and transitivity. These are the same as for events in Minkowski spacetime. $\mathcal{A}(x,\ y)$ is called an Alexandrov set of the elements $x$ and $y$ or a causal interval or an order interval. In Minkowski spacetime, an Alexandrov set of any pair of events is an empty set or a set of continuum. The local finiteness means that an Alexandrov set of any elements is finite. The physical meaning of this binary relation $\prec$ is causal or chronological order. By assumption a causal set describes spacetime and matter on a microscopic level. In the considered model, the set of vertexes and the set of edges are causal sets. A causal set approach to quantum gravity has been introduced by G. 't Hooft \cite{'t Hooft} and J. Myrheim \cite{Myrheim} in 1978. There are reviews of a causal set program \cite{Sorkin2005,Dowker2006,Henson2009,Wallden2010}.

The goal of the considered model is to describe physical objects as some self-organized structures of the graph. This self-organization must be the consequence of dynamics.

\section{Sequential growth dynamics}
\begin{figure}[ht]
	\centering	
		\includegraphics[trim=8cm 20cm 8cm 6cm]{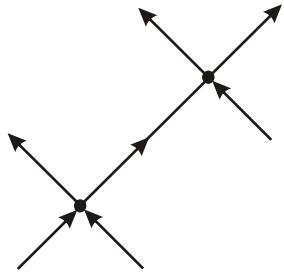}
	\caption{An example of the graph with 2 vertexes.}
	\label{fig:fig2}
\end{figure}

The model of the universe is an infinite graph. But any observer can only know finite graph. In a graph theory, by definition, an edge is a relation of two vertexes. Consequently some vertexes of finite graph have less than four incident edges. These vertexes have free valences instead the absent edges. These free valences are called external edges as external lines in Feynman diagrams. They are figured as edges that are incident to only one vertex. There are two types of external edges: incoming external edges and outgoing external edges. An example of the graph with 2 vertexes (Fig.\ \ref{fig:fig2}) possesses 1 edge, 3 incoming external edges, and 3 outgoing external edges. We can prove that the number of incoming external edges is equal to the number of outgoing external edges for any such graph \cite{1008.5169}\footnote{It should be noted that a set of halves of edges is considered in papers \cite{1008.5169,1004.5077,1106.6269}. The halves of edges as basic objects is introduced by D. Finkelstein G. McCollum and in 1975 \cite{FinkMcC1}. By some reasons, it is convenient to break the edge into two halves of which the edge is regarded as composed. The set of halves of edges in papers \cite{1008.5169,1004.5077,1106.6269} is isomorphic to the considered graph.}.

A finite graph is a model of a part of some process. The task is to predict the future of the process or to reconstruct the past. We can reconstruct the graph step by step. The minimal part is a vertex. We start from some given graph and add new vertexes one by one. This procedure is proposed in papers of author \cite{Krugly1998, Krugly2002}. Similar procedure and the term `a classical sequential growth dynamics' is proposed by D. P. Rideout and R. D. Sorkin \cite{RideoutSorkin} for other model of causal set dynamics.

We can add a new vertex to external edges. This procedure is called an elementary extension. There are four types of elementary extensions \cite{1004.5077}. There are two types of elementary extensions to outgoing external edges (Fig.\ \ref{fig:fig3} and\ \ref{fig:fig4}). This is a reconstruction of the future of the process.  In this and following figures the graph $\mathcal{G}$ is represented by a rectangle because it can have an arbitrary structure. The edges that take part in the elementary extension are figured by bold arrows. First type is an elementary extension to two outgoing external edges (Fig.\ \ref{fig:fig3}). The number $n$ of incoming or outgoing external edges is not changed by this elementary extension.
\begin{figure}
	\centering	
		\includegraphics[width=4cm,trim=8cm 16cm 8cm 5cm]{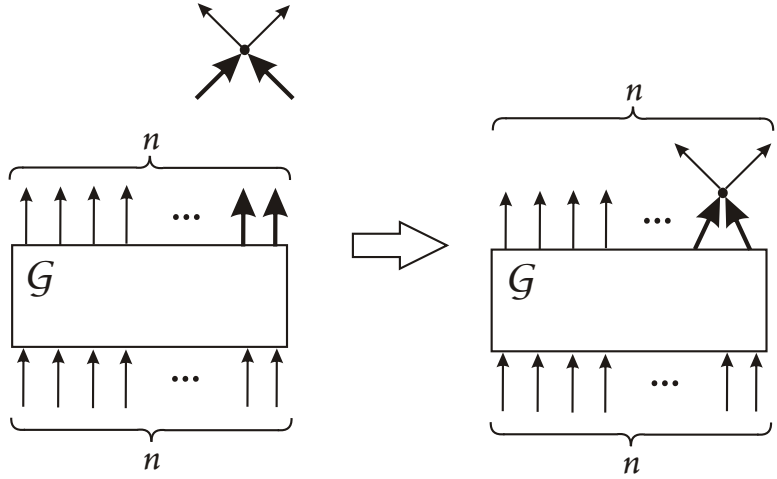}
	\caption{The first type of an elementary extension.}
	\label{fig:fig3}
\end{figure}
Second type is an elementary extension to one outgoing external edge (Fig.\ \ref{fig:fig4}). The numbers $n$ of  incoming external edges and outgoing external edges have increased by 1.
\begin{figure}
	\centering	
		\includegraphics[width=4cm,trim=8cm 16cm 8cm 5cm]{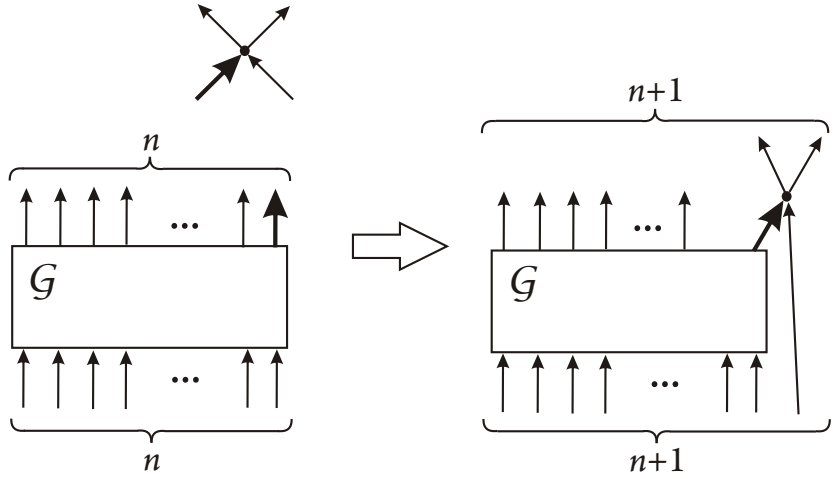}
	\caption{The second type of an elementary extension.}
	\label{fig:fig4}
\end{figure}
Similarly, there are two types of elementary extensions to incoming external edges (Fig.\ \ref{fig:fig5} and\ \ref{fig:fig6}). These elementary extensions reconstruct the past evolution of the process. Third type is an elementary extension to two incoming external edges (Fig.\ \ref{fig:fig5}). The number $n$ of incoming or outgoing external edges are not changed by this elementary extension.
\begin{figure}
	\centering	
		\includegraphics[width=4cm,trim=8cm 14cm 8cm 7cm]{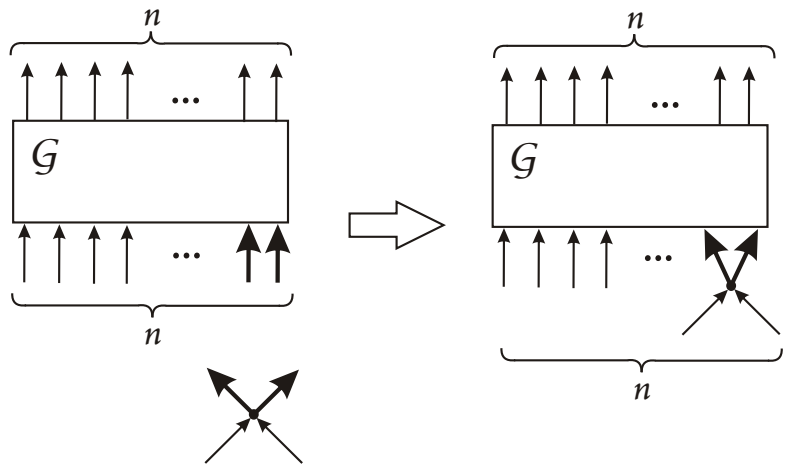}
	\caption{The third type of an elementary extension.}
	\label{fig:fig5}
\end{figure}
Fourth type is an elementary extension to one incoming external edge (Fig.\ \ref{fig:fig6}). The numbers $n$ of incoming external edges and outgoing external edges have increased by 1.
\begin{figure}
	\centering	
		\includegraphics[width=4cm,trim=8cm 14cm 8cm 7cm]{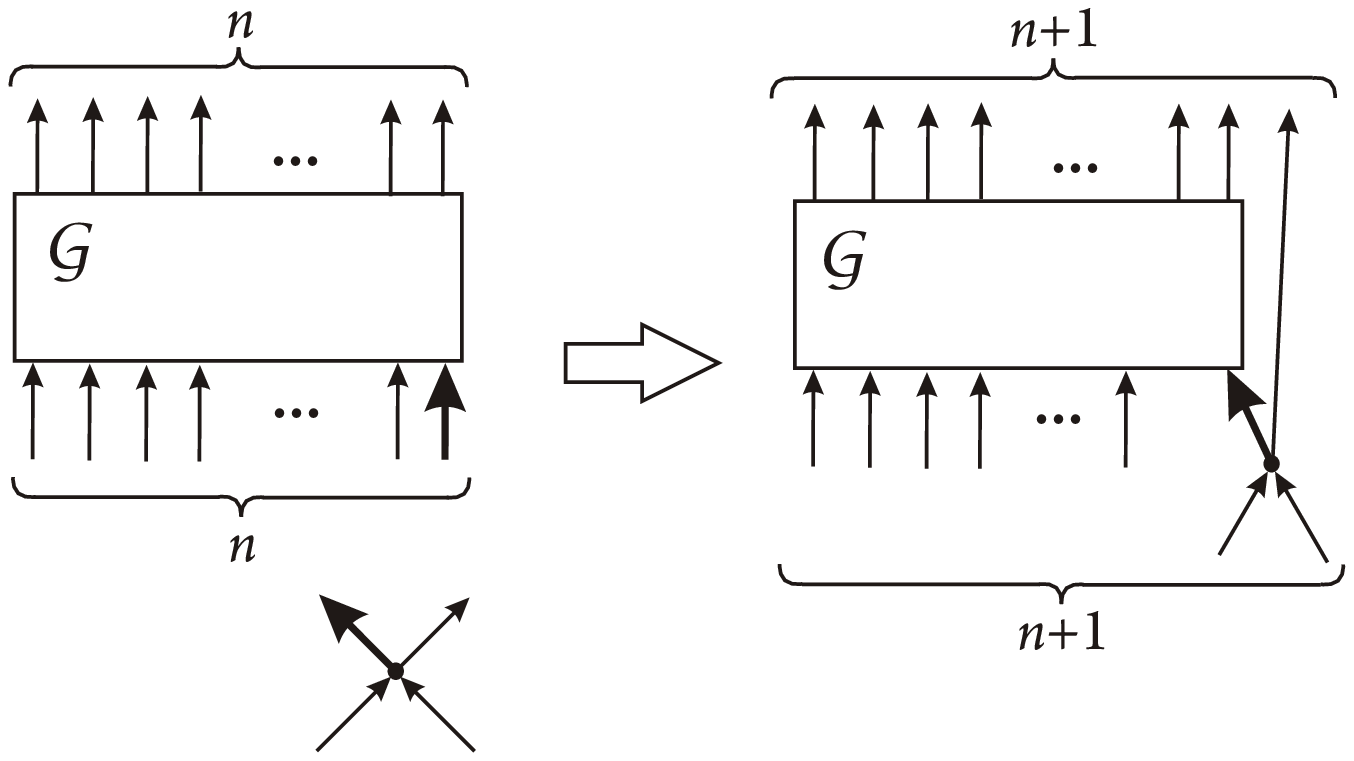}
	\caption{The fourth type of an elementary extension.}
	\label{fig:fig6}
\end{figure}
We can prove that we can get every connected graph by a sequence of elementary extensions of these four types \cite[Teorem~2]{1008.5169}.

By assumption, the dynamics of this model is a stochastic dynamics. We can only calculate probabilities of different variants of elementary extensions.

\section{The dynamics of binary alternatives}
Consider the dynamics that is based on binary alternatives. This alternative has two outcomes with probabilities $1/2$. This process has 1 bit of information. A binary alternative is considered as some primordial entities. This idea is considered in a set of papers. For example I introduce two citations of  C. F. von Weizs\"acker. \flqq It is certainly possible to decide any large alternative step by step in binary alternatives\frqq \ \cite[p. 222]{Weiz1}. \flqq $\dots$ the decision of an elementary binary alternative is the elementary process and hence the elementary interaction\frqq \ \cite[p. 94]{Weiz2}. Binary alternatives are discussed in the book ``Gravitation'' \cite[section~44.5]{gr} and other papers \cite{FinkMcC1, FinkMcC2, Weiz3}). This list of references is by no means complete. There is a recent paper of  M. Kober \cite{MK}. 

\begin{figure}
	\centering	
		\includegraphics[width=4cm,trim=8cm 15cm 8cm 6cm]{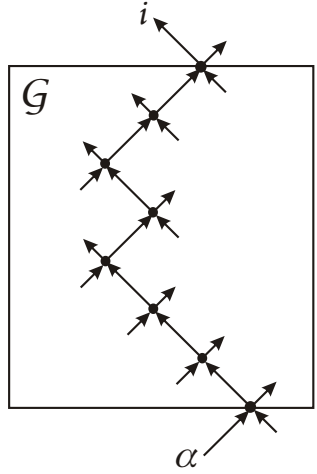}
	\caption{A choice of a directed path is a sequence of binary alternatives.}
	\label{fig:fig7}
\end{figure}

In the considered model a binary alternative is identified with an x-structure. Consequently the graph is a net of binary alternatives. Consider a directed path. Number outgoing external edges by Latin indices. Number incoming external edges by Greek indices. Latin and Greek indices range from 1 to $n$, where $n$ is the number of outgoing or incoming external edges. If we choose a directed path from any incoming external edge number $\alpha$, we must choose one of two edges in each vertex (Fig.\ \ref{fig:fig7}). Assume the equal probabilities for both outcomes independently of the structure of the graph. Then this probability is equal to $1/2$. This is the binary alternative. Consequently if a directed path includes $k$ vertexes, the choice of this path has the probability $2^{-k}$. We have the same choice for an opposite directed path.

Introduce an amplitude $a_{i\alpha }$ of causal connection of the outgoing external edge number $i$ and the incoming external edge number $\alpha$. By definition, put
\begin{equation}
\label{eq:1} a_{i\alpha}= a_{\alpha i}=\sum_{m=1}^M 2^{-k(m)}\textrm{,}
\end{equation}
where $M$ is the number of directed paths from the incoming external edge number $\alpha $ to the outgoing external edge number $i$ and $k(m)$ is the number of vertexes in the path number $m$. This definition has clear physical meaning. The causal connection of two edges is stronger if there are more directed paths between these edges and these paths are shorter.

Consider a following algorithm to calculate the probabilities of elementary extensions \cite{1106.6269}. There are three steps.

The first step is the choice of the elementary extension to the future or to the past. By definition, the probability of this choice is $1/2$ for both outcomes.

A new vertex is added to one or two external edges. The second step is the equiprobable choice of one external edge that takes part in the elementary extension. This is an outgoing external edge if we have chosen the future evolution in the first step. Otherwise this is an incoming external edge. The probability of this choice is $1/n$ for each outcome.

The third step is the choice of second external edge. Denote by $p_{ij}$ the probability to choose the outgoing external edge number $j$ if we have chosen the outgoing external edge number $i$ in the second step. By definition, put 
\begin{equation}
\label{eq:2} p_{ij}=\sum_{\alpha=1}^n a_{i\alpha} a_{\alpha j}\textrm{.}
\end{equation}

\begin{figure}[t]
	\centering	
		\includegraphics[width=4cm,trim=8cm 15cm 8cm 4cm]{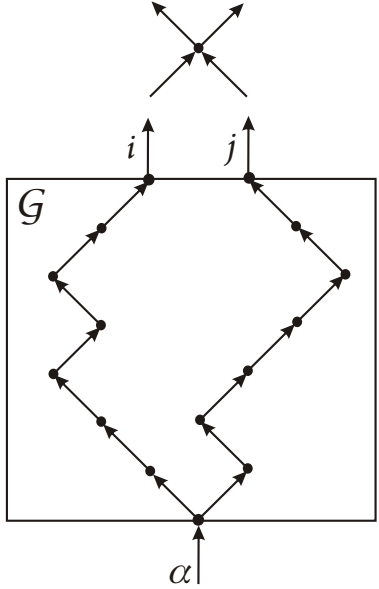}
	\caption{A new loop is generated by a new vertex.}
	\label{fig:fig8}
\end{figure}

Consider the meaning of this definition. The addition of a new vertex to two external edges forms a set of loops (Fig.\ \ref{fig:fig8}).
Each loop is formed by two directed paths. We can describe a loop as a product of these paths. The probability of the elementary extension is directly proportional to the sum of new loops that are generated by this elementary extension.

Similarly,
\begin{equation}
\label{eq:3} p_{\alpha \beta}=\sum_{i=1}^n a_{\alpha i} a_{i\beta}\textrm{,}
\end{equation}
where $p_{\alpha \beta}$ is the probability to add a new vertex to two incoming external edges numbers $\alpha$ and $\beta$.

The sum of all directed paths from any edge is equal to 1. We get the right normalization if we put the following definition for the probability to add a new vertex to one outgoing or incoming external edge, respectively.
\begin{equation}
\label{eq:4} p_{ii}=\sum_{\alpha=1}^n a_{i\alpha} a_{\alpha i}\textrm{,}
\end{equation}
\begin{equation}
\label{eq:5} p_{\alpha \alpha}=\sum_{i=1}^n a_{\alpha i} a_{i\alpha }\textrm{.}
\end{equation}

We can calculate the probability of any elementary extension by using this algorithm.

In this model, causality is defined as the order of vertexes and edges. But the causality has a real physical meaning only if the dynamics agrees with causality. The dynamical causality can be formulated in the following form. The probability to add a new vertex to the future can only depend on the subgraph that precedes this vertex \cite{RideoutSorkin}. Similarly, the probability to add a new vertex to the past can only depend on the subgraph that follows this vertex. 

The considered algorithm agrees with causality and has clear physical meaning. The probability to add a new vertex to two outgoing external edges is greater if their common past is larger, and this common past has the stronger connection with these edges.

\section{Physical objects}

\begin{figure}[t]
 \centering	
		\includegraphics[width=5cm,trim=8cm 16cm 8cm 7cm]{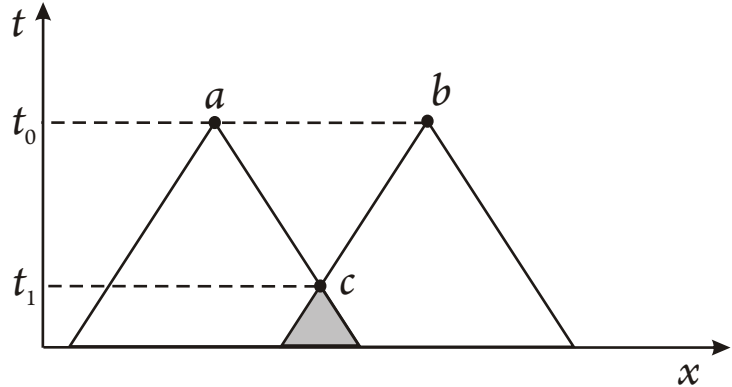}
 	\caption{The connection by the common past.}
	\label{fig:fig9}
\end{figure}

Usually in physical theory, physical objects are considered as primary entities in instants of time. Then a process is secondary, is a mapping of the objects or of their initial to their final states. An object has some state if it has some structure. But in relativity theory, structures cannot exist in instants of time. Consider a simple example (Fig.\ \ref{fig:fig9}). 
This is a 2-dimensional Minkowski space. Points $a$ and $b$ are simultaneous in the considered frame of reference. This instant of time is denoted by $t_0$. By definition, these points cannot have any connections in $t_0$. They can be connected only by the common past (the shaded triangle). The nearest point of this common past is the point $c$ in the instant of time $t_1$. The points $a$ and $b$ form a structure only together with $c$. This structure has a duration $t_0-t_1$. In the instant of time we have unconnected points. Any structure has finite duration. Consequently any structure is a process. \flqq According to relativity, the world is a collection of processes (events) with an unexpectedly unified causal or chronological structure. Then an object is secondary; is a long causal sequence of processes, a world line.\frqq\ \cite[p. 2923]{STC3}.

An antichain is a totally unordered subset of edges. Every two edges of this subset are not related by causal connection. A slice is an inextendible antichain. Every edge in the graph is either in the slice or causal connected to one of its edges. The set of all outgoing (or incoming) external edges is a slice. In the considered model, a slice of edges is a discrete analog of spacelike hypersurface. We can define objects (structures) in the slice of outgoing external edges by their common past. The connection of a pair of outgoing external edges numbers $i$ and $j$ ($i\ne j$) is defined by the probabilities $p_{ij}$. There is a scale hierarchy of the matter in the universe. Elements of the more deep level have stronger couplings. In the considered model, the strong coupling of the pair of the outgoing external edges numbers $i$ and $j$ is the high value of $p_{ij}$. This model is useful for numerical simulation. Now this investigation is started \cite{FPP6}. We start from 1 vertex and calculate 500 steps. There are many variants of the growth for a big graph. But usually there are very few variants with high probability. These are the couplings of outgoing external edges in the deepest level of objects.

We can define a threshold value $p_0$ for the deepest level of objects. If $p_{ij}\ge p_0$, the pair of the outgoing external edges numbers $i$ and $j$ belong to the same object of the deepest level. Let $p_{ij}$, where $i\ne j$ be elements of square matrix $\mathbf{p}(out)$ with zero main diagonal. Transform $\mathbf{p}(out)$. Replace $p_{ij}$ by 1 if $p_{ij}\ge p_0$. Replace $p_{ij}$ by 0 if $p_{ij}<p_0$. We get the matrix $\mathbf{s}(out)$. Consider $\mathbf{s}(out)$ as an adjacency matrix of some undirected graph $S(out)$. We have the isomorphism that takes each outgoing external edge to the vertex of $S(out)$. The vertexes numbers $i$ and $j$ of $S(out)$ are connected by an edge iff $p_{ij}\ge p_0$. By definition, an object of the deepest level is a connected subgraph of $S(out)$.  

A connected subgraph includes nonextendible cliques. By definition, a clique is a subgraph such that each pair of its vertexes is connected by an edge. In general case, a connected subgraph includes a set of overlapping and non-overlapping nonextendible cliques. These cliques form a frame of the object.

The cliques have the following property for the considered algorithm of sequential growth. Consider the addition of a new vertex to the outgoing external edges numbers $i$ and $j$. These two outgoing external edges become internal edges, and two new outgoing external edges appear. We must delete the vertexes numbers $i$ and $j$ of $S(out)$ and add two new vertexes. We get new graph $S_1(out)$. If the vertexes numbers $i$ and $j$ belong to the clique $C$ of $S(out)$, the two new vertexes belong to the clique $C_1$ of $S_1(out)$. We get the clique $C_1$ by deletion the vertexes numbers $i$ and $j$ and addition two new vertexes. Consequently the interior interactions in the frame of the object cannot destroy this frame.
 
Similarly, we can define objects in the slice of incoming external edges by their common future. We can define objects for arbitrary slice of edges. Consider two edges numbers $a$ and $b$ of this slice. Define an amplitude $a_{ia}$ of causal connection of the outgoing external edge number $i$ and the edge number $a$. Define an amplitude $a_{a\alpha}$ of causal connection of the edge number $a$ and the incoming external edge number $\alpha$. Consider an amplitude $p_{ab}$ of coupling of the edges number $a$ and $b$. By definition, put
\begin{equation}
\label{eq:6} p_{ab}=\sum_{\alpha=1}^n a_{a\alpha} a_{\alpha b}+\sum_{i=1}^n a_{ai} a_{ib}\textrm{.}
\end{equation}
If the edges numbers $a$ and $b$ are outgoing external edges, definition (\ref{eq:6}) coincides with (\ref{eq:2}). If the edges numbers $a$ and $b$ are incoming external edges, definition (\ref{eq:6}) coincides with (\ref{eq:3}). 

\section{Conclusion}

I hope that the existence of the small quantity of preferable variants of the growth is a symptom of self-organization. It is necessary to develop the methods to detect and analyze repetitive symmetrical self-organized structures during the numerical simulation of the sequential growth. One method is to consider the evolution of the graph $S(out)$ during the sequential growth. We can search the connected subgraphs of $S(out)$ by the algorithm of depth-first search (DFS). We can search the cliques of these connected subgraphs by the Bron-Kerbosh algorithm \cite{BK}. This is the task for further investigation.

I am grateful to Alexandr V. Kaganov and Vladimir V. Kassandrov for discussions and Ivan V. Stepanian for collaboration in numerical simulations.

\end{document}